# CONCISE: Compressed 'n' Composable Integer Set

Alessandro Colantonio[*,a], Roberto Di Pietro[a]

[a] *Università di Roma Tre, Dipartimento di Matematica, Roma, Italy*

**Abstract**

Bit arrays, or *bitmaps*, are used to significantly speed up set operations in several areas, such as data warehousing, information retrieval, and data mining, to cite a few. However, bitmaps usually use a large storage space, thus requiring compression. Nevertheless, there is a space-time tradeoff among compression schemes. The *Word Aligned Hybrid* (WAH) bitmap compression trades some space to allow for bitwise operations without first decompressing bitmaps. WAH has been recognized as the most efficient scheme in terms of computation time. In this paper we present CONCISE (**Co***mpressed* **'n'** **C***omposable* **I***nteger* **Se***t*), a new scheme that enjoys significatively better performances than those of WAH. In particular, when compared to WAH, our algorithm is able to reduce the required memory up to 50%, by having similar or better performance in terms of computation time. Further, we show that CONCISE can be efficiently used to manipulate bitmaps representing sets of integral numbers in lieu of well-known data structures such as arrays, lists, hashtables, and self-balancing binary search trees. Extensive experiments over synthetic data show the effectiveness of our approach.

## 1. Introduction

The term *bit array* or *bitmap* usually refers to an array data structure which stores individual bits. The main reason for adopting bitmaps is represented by their effectiveness at exploiting bit-level parallelism in hardware to perform operations quickly. A typical bit array stores $k \times w$ bits, where $w$ is the *word* size, that is the number of bits that the given CPU is able to manipulate via bitwise instructions (typically 32 or 64 in modern architectures), and $k$ is some non-negative integer. Bitmaps made up of $n$ bits can be used to implement a simple data structure for the storage of any subset of $\{1, 2, \ldots, n\}$. Specifically, if the $i$-th bit of a bitmap is "1" then the integer $i$ is within the integer set, whereas a "0" bit indicates that the number corresponding to its position is not in the set. This set data structure uses $\lceil n/w \rceil$ words of space—padding with zeros is a usual choice. Whether the *least significant bit* (the "rightmost" in the typical bitmap representation) or the *most significant bit* (the "leftmost" one) indicates the smallest-index number is largely irrelevant, but the former tends to be preferred and will be adopted throughout all the examples proposed in this paper. Therefore, if we want to compute the intersection or the union over integer sets represented by bitmaps of length $n$, we require $\lceil n/w \rceil$ bitwise AND/OR operations each.

Because of their property of leveraging bit-level parallelism, bitmaps often outperform many other data structures (e.g., self-balancing binary search trees, hash tables, or simple arrays or linked lists of the entries) on practical data sets, even those which are more efficient asymptotically. However, bitmaps show some drawbacks as well: they are wasteful in both time and space when representing very sparse sets—that is, sets containing few elements compared to their range. For such applications, *compressed bitmaps* should be considered instead. Classical compression algorithms introduce a computation overhead that may limit the benefits of using bitmaps. For example, well-known algorithms such as DEFLATE [1] effectively reduce the memory footprint, but performing set operations requires data to be decompressed, hence drastically increasing the computation time [2]. That is why compression schemes that allows for bitwise operations without first decompressing bitmaps are to be preferred, at the cost of having a memory footprint higher than other compression schemes. In this scenario, the *Word Aligned Hybrid* (WAH) bitmap compression algorithm is currently recognized as the most efficient one, mainly from a computational perspective [2]. It has been first proposed to compress bitmap indices of DBMS, but subsequent applications include visual analytics [3] and data mining [4], to cite a few. It uses a *run-length* encoding, where long sequences of 0's or 1's require a reduced number of bits by only storing the length of the sequences in place of the corresponding bit strings. WAH allows for set operations to be efficiently performed over the compressed representation. This property is guaranteed by the alignment with the machine word size. Figure 1 graphically explains what "alignment" means. Without loss of generality, suppose that words are made up of 32 bits. First, we logically partition the bitmap to compress into blocks of 31 bits, namely the word size minus one. In turn, sequences of consecutive 31-bit blocks containing all 0's or all 1's are being identified. The

*Corresponding author

*Email addresses:* colanton@mat.uniroma3.it (Alessandro Colantonio), dipietro@mat.uniroma3.it (Roberto Di Pietro)

*URL:* http://ricerca.mat.uniroma3.it/users/colanton/ (Alessandro Colantonio), http://ricerca.mat.uniroma3.it/users/dipietro/ (Roberto Di Pietro)

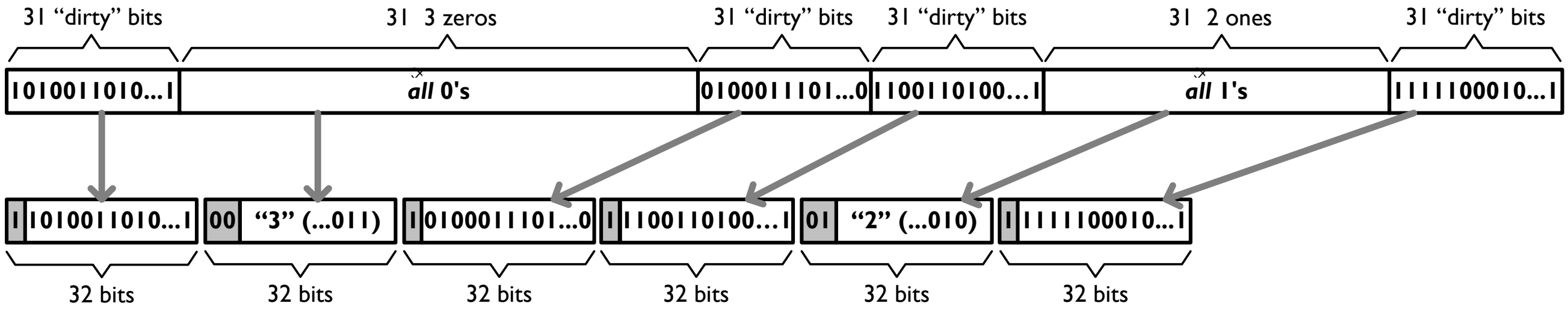

Figure 1: Word-aligned run-length encoding

compressed form is created as follows: if a 31-bit block contains both 0's and 1's, it is stored in a 32-bit word referred to as *literal* word, where the leftmost bit is set to 1. Otherwise, sequence of homogeneous 31-bit blocks of 0's or 1's are stored in a single 32-bit word referred to as *fill* word, where the first (leftmost) bit is 0, the second bit indicates the fill type (all 0's or all 1's) and the remaining 30 bits are used to store the number of 31-bit blocks.[1] Bitwise operations are fast because performing AND/OR over compressed bitmaps corresponds to performing AND/OR over 32-bit literal pairs (with just one CPU instruction), while sequences can be easily managed due to the same block granularity of literals.

This paper proposes a new compression scheme, CONCISE (**Co**mpressed *'**n**'* **C**omposable **I**nteger **Se**t), that outperforms WAH by reducing the size of the compressed bitmaps up to 50%, without affecting the performance of bitwise operations. For some widespread data configurations, CONCISE is even faster than WAH. CONCISE is based on the observation that very sparse bitmaps (that is, when there are few set bits followed by long sequence of unset bits) can be further compressed with respect to the WAH approach. Specifically, if $n$ is the number of bits that equals 1 in a sparse uncompressed bitmap, a WAH-based compressed bitmap requires about $2n$ words: one word for the literal that contains the set bit, and the other word for a literal that counts the subsequent unset bits. In fact, $2n$ is the upper bound[2] for the size of bitmaps with $n$ set bits [2, 5]. To achieve a better compression ratio on sparse bitmaps, CONCISE introduces the concept of "mixed" fill word. In particular, we allow to store the sequence length and the literal word made up of only one set bit within a single word of 32 bits. This reduces the worst case memory footprint to $n$ words. Since $n$ words is the minimum amount of memory required to represent $n$ integral values with classical data structures (e.g., with an array), CONCISE is always more efficient than other structures in terms of memory footprint. As for computation time, we show that it also outperforms efficient data structures such as hashtables or self-balancing binary search trees on set operations.

The remainder of the paper is organized as follows. The following section offers a detailed description of the CONCISE algorithm. The benefits of the proposed algorithm are then experimentally demonstrated in Section 3. Finally, Section 4 provides concluding remarks.

## 2. CONCISE Algorithm

Figure 2 shows an example of CONCISE-compressed bitmap made up of 5 words. Words #0, #3, and #5 are *literal* words where, similar to WAH, the leftmost bit indicates the block type ('1'), while the remaining bits are used to represent an uncompressed 31-bit block. Words #1, #2, and #4 are fill words: the first (leftmost) bit is the block type ('0'), the second bit is the fill type (a sequence of 0's or 1's), the following 5 bits are the *position* of a "flipped" bit within the first 31-bit block of the fill, and the remaining 25 bits count the number of 31-blocks that compose the fill minus one. When position bits equals 0 (binary '00000'), the word is a "pure" fill, similar to that of WAH. Otherwise, position bits indicate the bit to switch (from 0 to 1 in a sequence of 0's, or from 1 to 0 in a sequence of 1's) within the first 31-bit block of the sequence represented by the fill word. That is, 1 (binary '00001') indicates that we have to flip the rightmost bit, while 31 (binary '11111') indicates that we have to flip the leftmost one. If we consider bitmaps as a representation of integer sets, in Figure 2 Words #2 indicates that integers in the range 94–1022 are missing, but 93 is in the set since position bits say that the first number of the "missing numbers" sequence is an exception.

This approach allows to greatly improve the compression ratio in the worst case. Indeed, instead of having $2n$ words in the WAH-compressed form of $n$ sparse integers, we only require $n$ words in the CONCISE-compressed form for the same integer set. This way, CONCISE bitmaps always require less amount of memory than WAH bitmaps. As for performance, in the next section we will show that our proposal not only do not increase the computation time, but in some cases can also speed up operations thanks to the reduced number of words to manipulate. Since we have 25 bit for representing the length (minus one) of sequences of homogeneous bits, the maximum representable integer is $31 \times 2^{25} + 30 = 1\,040\,187\,422 \approx 2^{30}$, that is half of the positive integers that can be represented in a two's complement representation over 32 bits.

Due to space limitation, in Figure 3 we only describe with pseudo-code the three main parts of the proposed implementation of the CONCISE scheme. These parts are sufficient to have a complete comprehension of the algorithm. However, the complete source code is available on *SourceForge* since January 2010. The CONCISE algorithm has been coded in Java. More details about the actual implementation of CONCISE, as well as the code used for the comparative analysis, can be found at `http://sourceforge.net/projects/concise`.[3] In Figure 3 the following notation has been adopted:

- "|" indicates the bitwise OR operator, "&" the bitwise AND, "∼" the bitwise NOT, and "≪" means the left-shift operator.
- 80000000h is an instance of a 32-bit word expressed in hexadecimal notation. That is, 80000000h indicates a 32-bit word with the highest-order (leftmost) bit set to 1 and all other bits set to 0.

[1] In the paper of Wu et al. [2], the most significant bit is complemented with respect to the example of Figure 1, that is literals start with 0 and fills start with 1. Though this does not change the semantic of the approach, we use the configuration of Figure 1 since it reflects the proposed implementation of CONCISE.

[2] Apart from a few additional data required to manage the actual coding of the algorithm, such as the number of the leftover bits within the last literal word.

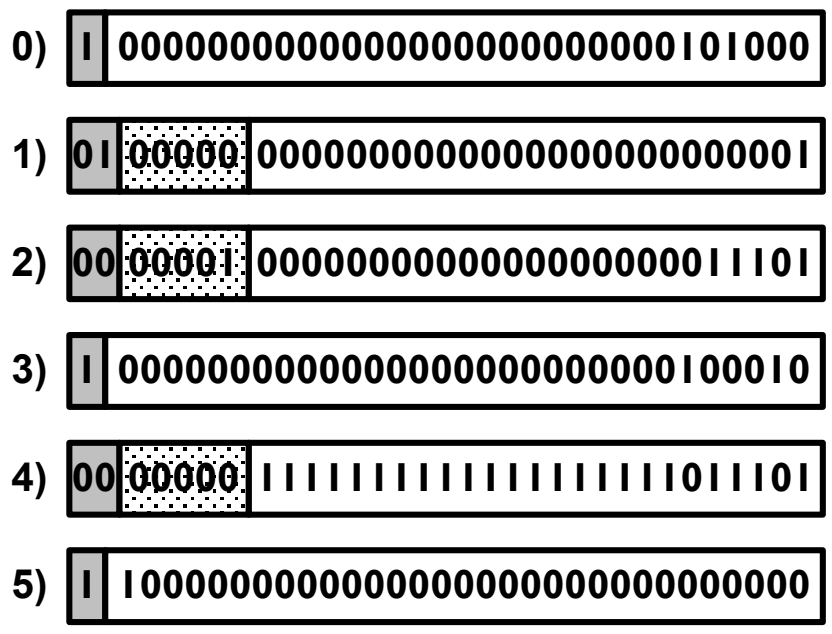

Figure 2: Compressed representation of the set {3, 5, 31–93, 1024, 1028, 1 040 187 422}. The word #0 is used to represent integers in the range 0–30, word #1 for integers in 31–92, word #2 for integers 93–1022, word #3 for integers 1023–1053, word #4 for integers 1054–1 040 187 391, and word #5 for integers 1 040 187 392–1 040 187 422.

We also requires additional bit operations that can be efficiently performed in most machines. In particular:

- BITCOUNT($n$) counts the total number of set bits in the binary representation of the specified integral value $n$—also known as *population count* or *Hamming weight*. It is natively provided by some processors via a single instruction to calculate it (i.e., `POPCNT` of Intel SSE4 instruction set [6]). For processors lacking this feature, there are efficient branch-free algorithms which can compute the number of bits in a word using a series of simple bit operations [7].
- TRAILINGZEROS($n$) counts the number of 0's following the lowest-order (rightmost) bit in the binary representation of the specified integral value.[4]
- CONTAINSONEBIT($n$) checks whether the given number contains only one set bit, and it can be efficiently performed by computing $n \,\&\, (n - 1) = 0$.
- $n \times 31$ can be performed by doing $(n \ll 5) - n$. Several bit hacks exist to efficiently compute $n$ mod 31.

## 3. Algorithm Analysis

In this section we report on the results of a comparative analysis among our CONCISE implementation, WAH, and some classical data structures to manage integer sets. The testbed was represented by a notebook equipped with an Intel Core™2 Duo CPU P8600 at 2.40 GHz and 3GB of RAM. All algorithms were coded in Java SE6. Since we did not find any reliable implementation of WAH in Java, testing WAH was performed by "switching off" the possibility of having "mixed" fill words in our implementation of CONCISE. This also assures that differences in performances are mainly due to the compression schema, and not to the given implementation. As for other data structures, we used the classes provided by the Java package java.util. Pseudo-random numbers were generated through the algorithm described in [8] due to its provable good uniform distribution and very large period.

Table 1 reports some characteristic about the memory footprint of the data structure under analysis. For each structure, we report the number of bytes required to store each integral number, whether the structure allows for duplicate elements, and if the items are kept sorted or not. CONCISE is the more efficient data structure in terms of memory occupation. In fact, classical structures incur an additional linear space overhead for pointers, while WAH requires twice the memory of CONCISE when both algorithms are able to compress data—that is, in presence of sparse datasets.

Figure 4 reports on experimental time-space analysis results. In our experiments, we generated sets of $10^5$ integers, ranging from 0 to a variable maximum value max. Within this range, numbers were generated at random according to two different probability distributions: *uniform* and *Zipfian*. In particular, at each generation of a pseudo-random number $a \in [0, 1)$, in uniform sets an integer corresponding to $\lfloor a \times \mathsf{max} \rfloor$ was added, where $\mathsf{max} = 10^5/d$ by varying $d$ (the "density") from 0.005 to 0.999. Numbers were generated till reaching the cardinality of $10^5$. Similarly, in Zipfian sets, at each number generation, an integer corresponding to $\lfloor \mathsf{max} \times a^4 \rfloor$ was added, where $\mathsf{max} \in [1.2 \times 10^5,\ 6 \times 10^9]$. In this way, we generated skewed data such that most of the integers were concentrated to lower values, while numbers with high values were very sparse. The reason for a Zipfian distribution is that, according to the Zipf's

[3] The first release can be downloaded at `http://concise.svn.sourceforge.net/viewvc/concise?view=rev&revision=1`. Following versions can present some improvements, hence being slightly different from the algorithm described in this paper.

[4] For more details about bit hacks, see, for example, `http://graphics.stanford.edu/~seander/bithacks.html`.

```
procedure Append(words[·], top, max, i)
    {first append}
    if words[·] is empty then
        f ← ⌊i/31⌋
        if f = 0 then
            top ← 0
        else if f = 1 then
            top ← 1
            words[0] ← 80000000h {literal}
        else
            top ← 1
            words[0] ← f − 1 {fill}
        end if
        words[top] ← 80000000h | (1 ≪ (i mod 31))
        max ← i
        return words[·], top, max
    end if

    b ← i − max + (max mod 31) {position of next bit to set}

    {check if zeros are required before the new word}
    if b ≥ 31 then
        f ← ⌊b/31⌋
        if f = 0 then
            top ← top + 1 {just add a new word to set the bit}
        else
            {add a 0's before the new word}
            if ContainsOneBit(words[top]) then
                {merge the previous word}
                words[top] ← (1 + TrailingZeros(words[top])) ≪ 25 | f
                top ← top + 1
            else
                top ← top + 2
                if f = 1 then
                    words[top − 1] ← 80000000h {literal}
                else
                    words[top − 1] ← f − 1 {fill}
                end if
            end if
        end if
        {prepare the new word}
        b ← b mod 31
        words[top] ← 80000000h
    end if

    words[top] ← words[top] | (1 ≪ b) {set the bit}
    max ← i
    Compress(words[·], top)
    return words[·], top, max
end procedure
```

(a) Append of a new integer $i$ that is greater than the maximal appended integer max. It checks whether the bit to set is within the last literal, or if we need to add a sequence of 0's before setting the bit. It also returns top, namely the index of the last word of words.

```
procedure Compress(words[·], top)
    if top = 0 then {check if the set is empty}
        return words[·], top
    end if

    φ0 ← words[top] = 80000000h {last word all 0's}
    φ1 ← words[top] = FFFFFFFFh {last word all 1's}
    if ¬φ0 ∧ ¬φ1 then {compress only if there are all 0's or all 1's}
        return words[·], top
    end if

    φ'0 ← words[top − 1] & C0000000h = 00000000h {0's fill}
    φ'1 ← words[top − 1] & C0000000h = 40000000h {1's fill}
    if (φ0 ∧ φ'0) ∨ (φ1 ∧ φ'1) then {previous word is the same fill}
        top ← top − 1
        words[top] ← words[top] + 1
        return words[·], top
    end if

    if (φ0 ∧ φ'1) ∨ (φ1 ∧ φ'0) then {previous word is a different fill}
        return words[·], top
    end if

    w ← words[top − 1]
    if φ1 then
        w ← ~w
    else
        w ← w & 7FFFFFFFh
    end if
    if w = 0 ∧ BitCount(w) = 1 then
        top ← top − 1
        if φ1 then
            words[top] ← 40000001h {1's fill}
        else
            words[top] ← 00000001h {0's fill}
        end if
        if BitCount(w) = 1 then {check dirty bit}
            words[top] ← words[top] | (1 + TrailingZeros(w)) ≪ 25
        end if
    end if
    return words[·], top
end procedure
```

(b) Compression algorithm. It tries to "merge" the literal that is in the last word of the compressed bitmap with the previous word.

```
procedure PerformOperation(S1, S2, ⋆)
    R.top ← 0
    while HasMore(S1.words[·]) ∧ HasMore(S2.words[·]) do
        R.top ← R.top + 1
        R.words[R.top] ←
                NextLiteral(S1.words[·]) ⋆ NextLiteral(S2.words[·])
        Compress(R.words[·], R.top)

        {check if we just created a fill}
        if R.words[R.top] & 80000000h = 00000000h then
            s ← min{Length(S1.words[·]), Length(S2.words[·])}
            if s > 0 then
                Skip(S1.words[·], s), Skip(S2.words[·], s)
                R.words[R.top] ← R.words[R.top] + s
            end if
        end if
    end while

    {copy remaining words}
    if ⋆ is bitwise OR or XOR then
        if HasMore(S1.words[·]) then
            append to R all the remaining words of S1
        else
            append to R all the remaining words of S2
        end if
    else if ⋆ is bitwise AND-NOT then
        append to r all the remaining words of S2
    end if
    return R
end procedure
```

(c) Bitwise operations between two compressed bitmaps. "⋆" indicates the desired binary operation. The algorithm iterates over both word arrays and performs the operation represented by ⋆. HasMore() indicates if we reached the last word, while NextLiteral() extracts the next literal. When applied to a fill, NextLiteral() returns all the literals represented by the fill: it stores the position of the current literal within a fill, making it possible to "skip" a specified number of blocks by calling Skip(). Length() gives the length of a fill represented by the word $w$ (i.e., $w$ & 01FFFFFFh), or 0 for literals.

Figure 3: The Concise algorithm. Figure (a) describes how to create new compressed bitmaps (indicated with the array words[·]) by "appending" integral numbers in ascending order—namely we can only add integrals that are greater than the last appended one. Figure (c) describes how to apply AND/OR/XOR/AND-NOT operations over compressed bitmaps. Finally, Figure (b) is used by both Append and PerformOperation to compress bits.

Table 1: Memory footprint analysis among standard Java structures from the package java.util, WAH, and CONCISE

| *Data Structure* | *Distinct?* | *Sorted?* | *Bytes/Item*[a] | *Explanation* |
|---|---|---|---|---|
| ArrayList | | | 4 | Simple array. It is the simplest data structure. It is internally represented by an array of pointers (4 bytes each) to Integer instances. |
| LinkedList | | | 24 | Linked list. Each element of the list requires $4\times3$ bytes (4 bytes to point the Integer instance, 4 bytes to point the previous element, and 4 bytes to point the next one), plus 8 bytes used by Java for each object instance, and 4 padding bytes. |
| HashSet | ✓ | | ≥30 | Hashtable. Each element requires $4 \times 4$ bytes (4 bytes to point the key—the Integer instance—, 4 bytes to point the value—not used in our tests—, 4 bytes to point the next table entry in case of collision, and 4 bytes to store the hash value of the key), plus 8 bytes internally used by Java for the table entry. Moreover, we require an array of pointers (4 bytes for each element), considering that a hashtable must be greater than the maximum number of allowed elements in order to reduce the number of collisions. |
| TreeSet | ✓ | ✓ | 32 | Self-balancing, red-black binary search tree. Each node of the tree requires $4 \times 5 + 1$ bytes (4 bytes to point the key—the Integer instance—, 4 bytes to point the value—not used in our tests—, 4 bytes to point the left node, 4 bytes to point the right node, 4 bytes to point the parent node, and 1 byte for the node color), plus 8 bytes internally used by Java for the node object, and 3 padding bytes. |
| BitSet | ✓ | ✓ | $1/8 \div 2^{28}$ | Uncompressed bitmap. Each integral value is stored in a bit. In the worst case, we need a long sequence of zeros and then a word to store the integral. If we only consider positive integral numbers represented in two's complement over 32 bits, the greatest number is $2^{31} - 1$. In this case, we need a bitmap of $2^{28}$ bytes. In the best case, all integers represent a sequence of consecutive numbers, thus requiring only 1 bit on average. |
| WAH | ✓ | ✓ | $\sim 0 \div 8$ | In the worst case, namely when numbers are very sparse, we need a literal word to store the integer (4 bytes) and a fill word to store a zero sequence (4 bytes). In the best case, all integers represent a sequence, thus requiring only 1 fill word (4 bytes) to represent all of them. |
| CONCISE | ✓ | ✓ | $\sim 0 \div 4$ | In the worst case, namely when numbers are very sparse, we store each integer in each mixed fill word (4 bytes). In the best case, all integers represents a sequence, thus requiring only 1 fill word (4 bytes) to represent all of them. |

[a]Please note that each Java object requires at least 8 + 8 bytes of memory: 8 bytes to represent information that are internally managed by the Java Virtual Machine (JVM), while user-defined object data should be aligned to a multiple of 8 bytes—in case, memory is padded with 0's. Moreover, in standard Java collections (namely any class implementing the Collection interface, such as ArrayList, LinkedList, HashSet, and TreeSet), integral numbers are stored within Integer instances. This means that each number to store requires 16 bytes (8 for internal JVM data, 4 for the integer, and 4 for padding) in addition to those reported in this table. Instead, BitSet, WAH, and CONCISE directly stores integers within an array of ints, hence avoiding to "waste" this additional space.

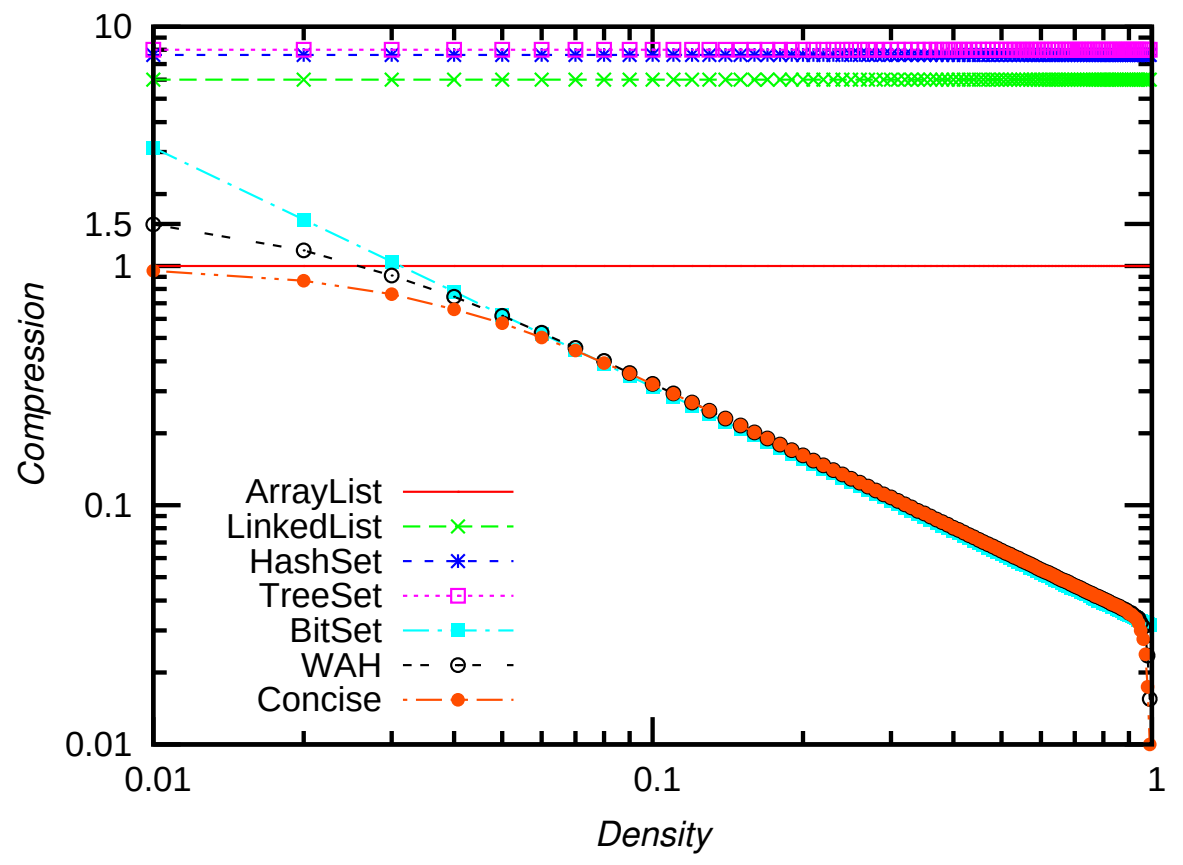

(a) Memory footprint in uniform distribution

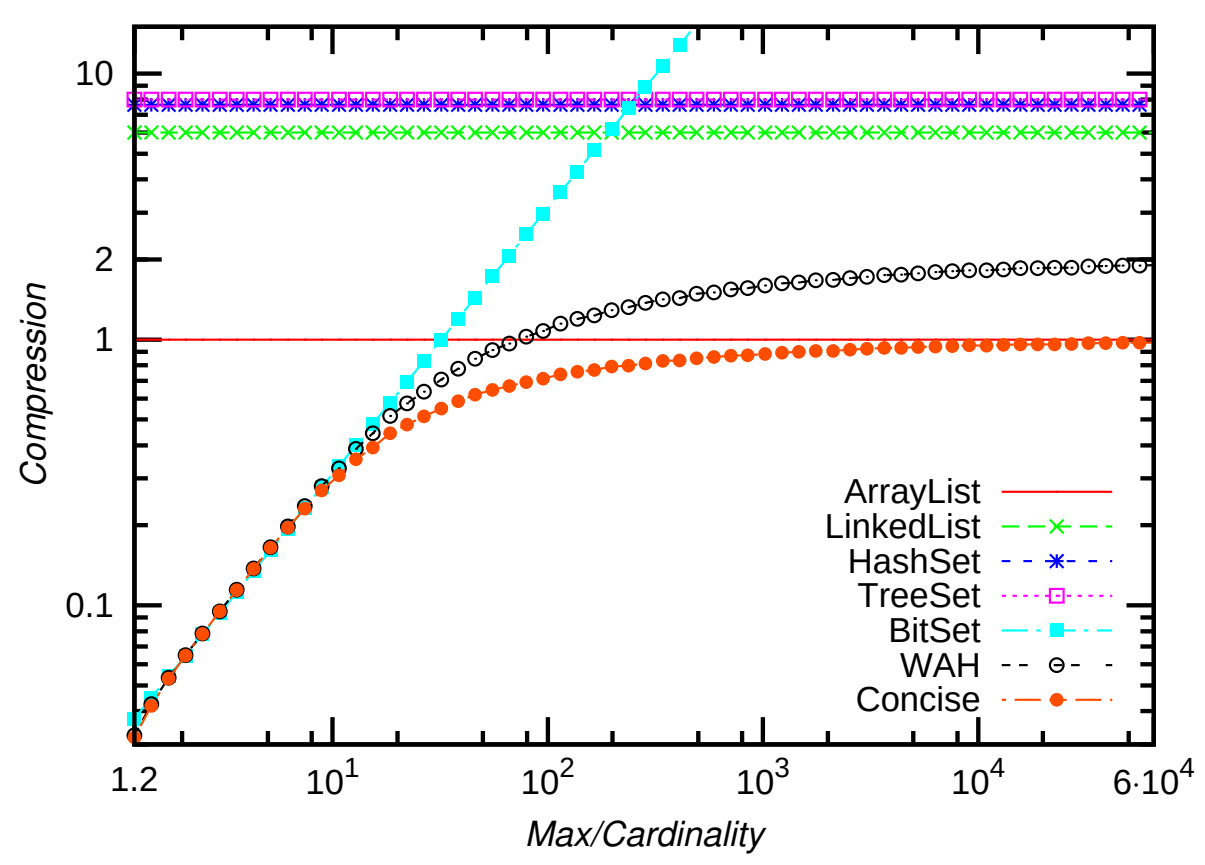

(b) Memory footprint in Zipfian distribution

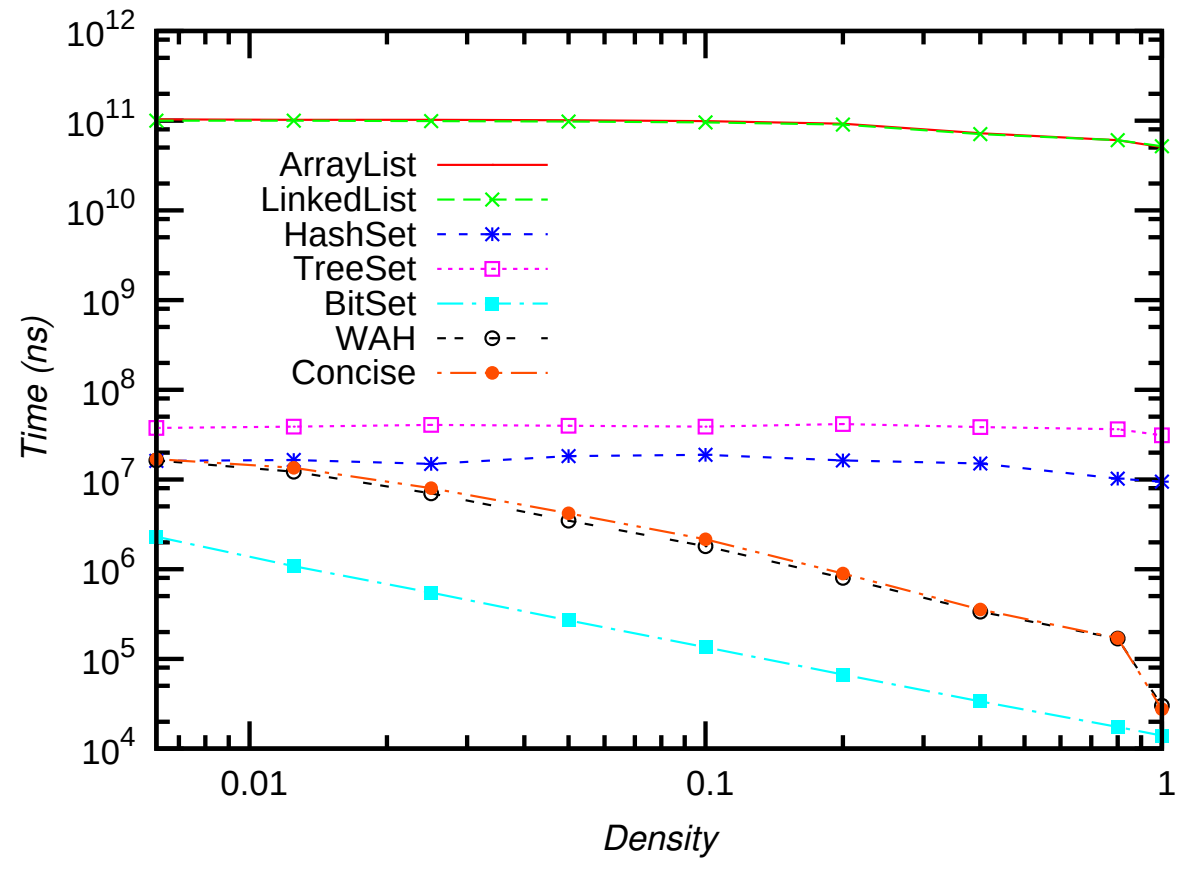

(c) Intersection in uniform distribution

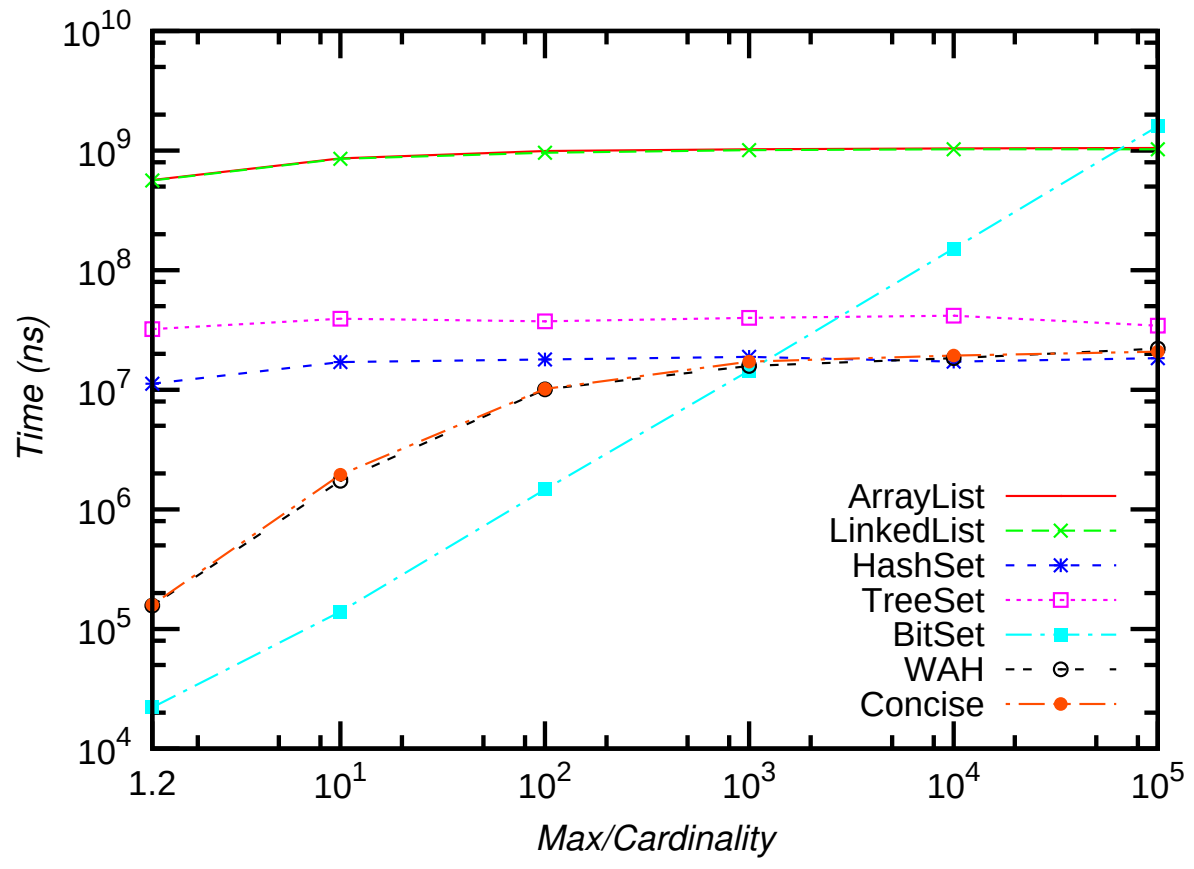

(d) Intersection in Zipfian distribution

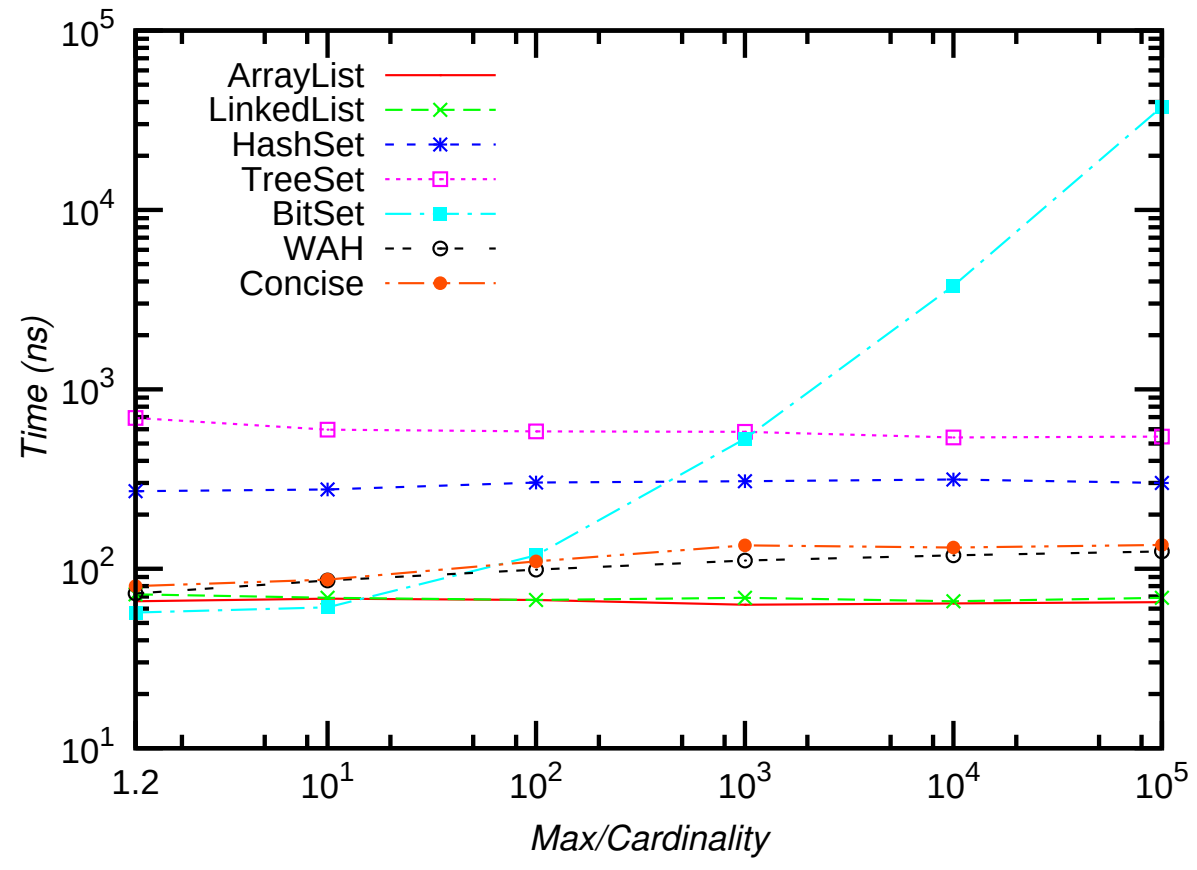

(e) Append in Zipfian distribution

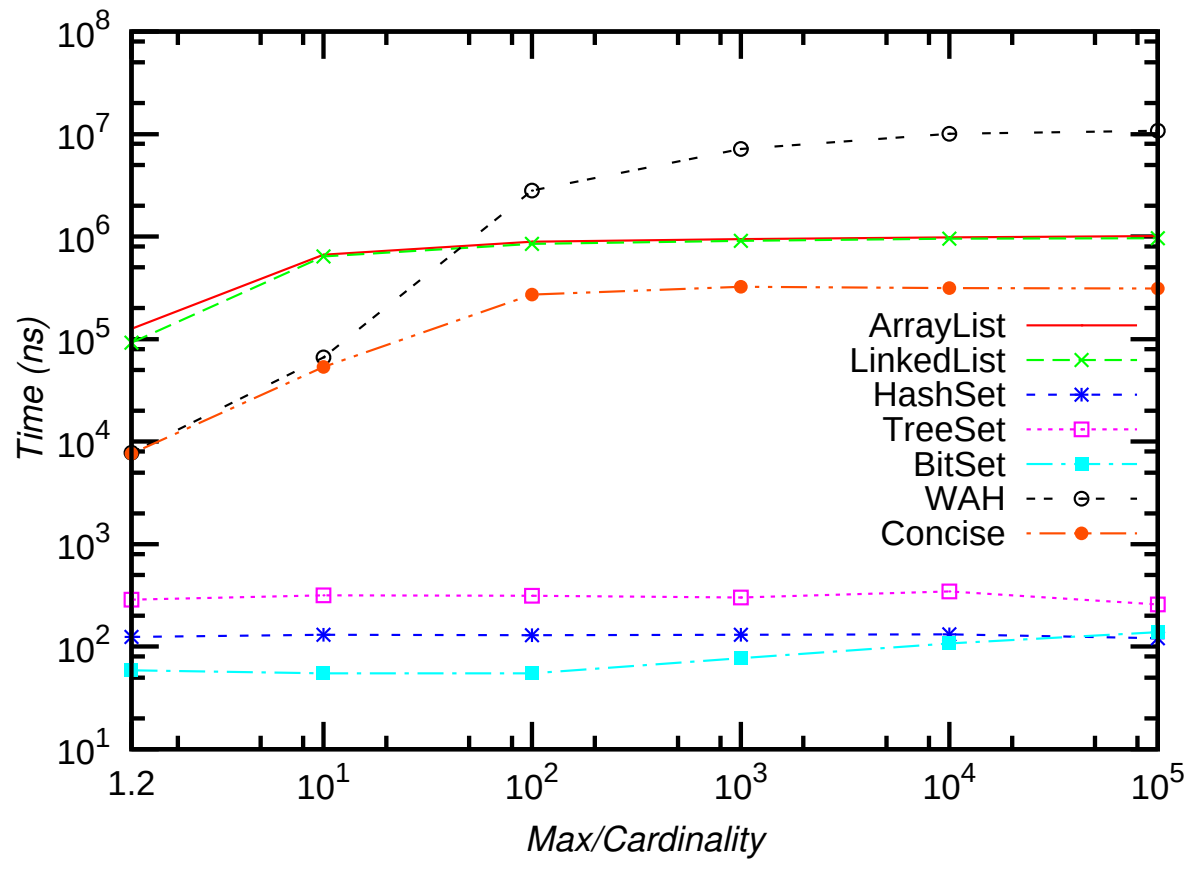

(f) Removal in Zipfian distribution

Figure 4: Time and memory measurements. "Compression" means the ratio between the number of 32-bit words required to represent the compressed bitmap and the cardinality of the integer set. "Density" means the ratio between the cardinality of the set and the number range. "Max/Cardinality" means the ratio between the maximal value (i.e., the number range) and the cardinality of the set—that is, the inverse of the density. Time measurement are expressed in nanoseconds. Since the experiments were performed in a multitasking environment, to have a more accurate time measurement, each experiment was performed 100 times, and the average reported.

law, many types of data studied in the physical and social sciences can be approximated with a Zipfian distribution [9].

Figure 4a reports on the memory occupation of one randomly generated set. It demonstrates that, according to [2], when density is below 0.05, WAH starts to compress. Since CONCISE is able to compress the *same* bitmaps that WAH can compress, both algorithms start to compress after the same density threshold. However, CONCISE always has a better compression ratio, which tends to be half of that of WAH when the density approaches zero. In Figure 4b, results are very similar, but it is more evident the compression ratio of WAH is twice of that of CONCISE as the data becomes more and more sparse. As expected, uncompressed bitmaps (BitSet) continue to increase as the maximum integer value grows, while other data structures are not affected by the data density.

Figure 4c reports on intersection time of two sets, namely the time required for the identification of shared numbers between two sets. We do not show results for union and set difference because they have demonstrated a very similar computation time. For Java classes, intersecting corresponds to calling Collection.retainAll() and BitSet.and() methods. Notice that WAH and CONCISE are always faster than Java structures, apart from BitSet that is far much faster when the set is not sparse. However, BitSet performance drastically decreases when data becomes very sparse. Again, Java data structures are not affected by the density. In our experiments, we also noted (as expected) that the intersection time changes linearly with respect to the cardinality of the set. Similar considerations can be done for Figure 4d. Notice that other curves can be justified in the following way: lists (ArrayList and LinkedList) requires a full set scan to perform the intersection, binary tree (TreeSet) a logarithmic time search and hashtable (HashSet) an almost constant time search of shared elements.

In turn, we analyzed the time to add single numbers to a set. In Figure 4e we report on the append time, namely on the addition of a new number that is strictly greater than the existing ones. Formally, $S \cup \{e\}$ when $\forall e' \in S : e' < e$. This corresponds to a sequence of calls to Java Collection.add() or BitSet.set() where numbers are first sorted. The append operation is the fastest way to add numbers to CONCISE and WAH bitmaps. Instead, sorting does not influence the performance of other data structures. Notice that the append time is constant for all structures and, as we observed in our experiments, it does not greatly change as cardinality grows. However, for very sparse data, the BitSet class spend most of its time in allocating memory, hence resulting in poor performances.

Finally, we analyzed the time to remove a single number from a set. In Figure 4f we indicate the corresponding execution time. Since both WAH and CONCISE do not explicitly support removal of single elements, we implemented it by performing the set difference between the given set and a singleton. Note that the same thing can be done for the addition of new integers when the append operation is not possible, by just performing a union between the set and a singleton. In this case, the reduced size of the compressed bitmap causes that CONCISE is much more faster than WAH on sparse datasets.

## 4. Conclusions

Because of their property of leveraging bit-level parallelism, computations over bitmaps often outperform computations over many other data structures such as self-balancing binary search trees, hash tables, or simple arrays. We demonstrated, through experiment on synthetic datasets, that bitmaps can be very efficient when data are dense.

However, when data become sparse, uncompressed bitmaps perform poorly due to the waste of memory. In this paper we introduced a new compression scheme for bitmaps, referred to as CONCISE, that is a good trade-off between the speed of uncompressed bitmaps and the required memory. Indeed, CONCISE outperformed all analyzed data structures in terms of memory occupation, as well as WAH, the best known compression algorithm that allows for set operations directly on the compressed form. As for computation time, CONCISE also outperformed classical data structures for set operations.

However, accessing individual elements can be expensive for both CONCISE and WAH. If random access is more common than sequential access, and the integer set is relatively small, classical data structures may be preferable.